\documentclass[10pt,tcilatex]{article}
\usepackage{amssymb}
\usepackage{amsmath}
\usepackage[dvips]{graphicx}
\usepackage{bm}

\input{tcilatex}
\begin{document}
\noindent Note di Matematica  \qquad\qquad\qquad\qquad\qquad\ ISSN 1123-2536, e-ISSN 1590-0932\\
Note Mat. 32 (2012) n. 1, 105-124.  \qquad\quad  doi:10.1285/i15900932v32n1p105\\

{\noindent \textbf{\Large Statics and dynamics of fluids in nanotubes}}%
\medskip \vskip 0,5 cm {\noindent\textbf{Henri Gouin}}\newline
{\noindent \emph{{\footnotesize M2P2,\ U.M.R. C.N.R.S. 7340} \& University of Aix-Marseille
\newline
Av. Escadrille Normandie-Niemen, 13397 Marseille Cedex 20, France.} }
\newline
{\noindent{\small \emph{E-mails: henri.gouin@univ-amu.fr; henri.gouin@yahoo.fr }}} \bigskip

{\noindent\textbf{Abstract.}}   The purpose of this article is to study the
statics and dynamics of nanotubes by using the methods of continuum mechanics.
 The nanotube
can be filled with only a liquid or a vapour phase according to the physicochemical characteristics of the wall and to the
disjoining pressure associated with the liquid and vapour mother bulks of the fluid,  regardless of the
nature of the external mother bulk. In dynamics,  flows through nanotubes can be much more
important than classical Poiseuille flows. When the external mother bulk is of
vapour, the flow can be a million times larger than the classical flows when slippage on wall does not exist. \newline


\newline

\section{Introduction}

On one hand, it is well documented in the modern literature that the
conventional laws of capillarity are not adequate when applied to fluids
confined by porous materials \cite{Bear}. On the other hand, the technical
development of sciences allows us to observe phenomena at length scales of a
very few number of nanometres. This nanophysics allows to infer
applications in numerous fields, including medicine and biology. Iijima
often cited as the discoverer of carbon nanotubes \cite{Iijima}, was
fascinated by Kr\"atschmer \emph{et al}' Nature paper \cite{Kr}, and decided to launch out into a
detailed study of nanomaterials. The recent applications revealed new
behaviors that are often surprising and essentially different from those
usually observed at macroscopic scale but also at microscopic scale \cite%
{Harris}. Nonetheless, simple models  proposing qualitative behaviors
need  to be developed in the different fields of nanosciences; our aim
is to investigate the fluid-solid interaction in static  as well
as in dynamic conditions by differential calculus in continuum mechanics.
\newline
As it was pointed out in experiments, the density of liquid water  changes
in narrow pores \cite{Ball}; an analytic asymptotic expression was obtained
with an approximation of London potentials for liquid-liquid and
solid-liquid interactions, which yields the surface interaction energy \cite%
{Gouin 2}. With the aim to propose an analytic expression of the density for
liquid films with a nanometer thickness near a solid wall, we add together the
interaction energy at the solid surface to a density-functional
at the liquid-vapour interface and a square-gradient functional which
represents the volume free energy of the fluid \cite{Gouin 1,Gouin 3}. The
obtained functional allows to get a differential equation and boundary condition which yield the
density profile in cylindrical nanotubes. \newline
For shallow water, the flows of liquids on solids are mainly represented by
using the Navier-Stokes equations associated with the adherence condition at
the walls \cite{oron}. Recent experiments in nanofluidics seem to prove that
the adherence condition is often disqualified \cite{De Gennes}. So, we can
draw consequences which differ from results of classical models; the model
we are presenting reveals an essential difference between the flows of
microfluidics and those of nanofluidics \cite{Tabeling}. The simple laws
of scales cannot be only taken into account. The film-solid interactions are
accounted for in terms of the disjoining pressure. This concept of
disjoining pressure has been introduced by Derjaguin in 1936  as the
difference between the    pressure in a phase adjacent to a surface
confining it  and the pressure in the bulk of this phase \cite{Derjaguin,
Gouin 6}. We have previously seen  that the gradient of thickness along
layers creates a gradient of disjoining pressure that induces driving forces
along the layer \cite{Gouin 7}. Moreover, we had noticed that the stability
criterion of the flow issued from the equation of motions fits with the
results of Derjaguin's school \cite{Gouin 7}. \newline
The liquid flows through nanotubes  also depend on the wetting conditions on the wall. Some
phase transitions can appear and drastically change the liquid flows through
nanotubes. Since fifteen years the literature is abundant about nanotube
technology and flows inside nanotubes \cite{Mattia}. \newline

The aim of this paper is not to redo the literature but to emphasis on
liquid compressibility near solid walls in nanoscale conditions. We consider a nanotube made up
of a cylindrical hollow tube whose diameter is of some nanometres. The length
of the nanotube is microscopic and the edge of the cylinder is a solid made
out of carbon or other materials \cite{Harris}. The nanotube is immersed in
a liquid or a vapour made up of the same fluid. The fluid
fills the interior of the nanotube. The fluid is modeled by a van der Waals
fluid \cite{Korteweg,van der Waals} for which the surdeformations are taken
into account (we called \emph{capillary fluid} or \emph{Cahn and Hilliard fluid}
\cite{Cahn,Gouin 1}). The volume free energy of the fluid is a function not only of the
density but also of the gradient of density. The  conditions on the wall take
account of the fluid density at its immediate proximity \cite{Gouin 2}.
We first recall the equations of equilibrium and motion of
capillary fluids \cite{Gouin 1}. These fluids can be modeled on the
interfaces as   the fluids in the immediate vicinity of solid media
\cite{Gouin 2}. In liquid or vapour phase, it is possible to
express the chemical potential with a development in a linear form taking
account of the isothermal sound velocity values in the bulks \cite{Gouin 6}.
The main expansions of the free energy in liquid and vapour phases are
deduced. In nanofluidics, the interactions between   fluid and
solid wall  dominate over the hydrodynamic behaviour of the fluid \cite{Ball}.
Boundary conditions are embedding effects; they are expressed thanks to a
surface energy associated with a molecular model in mean field theory \cite%
{Gouin 3}.

In the case of   cylindrical geometry, a differential equation with respect to the fluid density is obtained. Then,   the profile of
the fluid density in a cylinder can be deduced. The result is applied to nanotubes when the
diameter ranges from a little number of nanometres to one hundred
nanometres. Depending on the disjoining pressure between liquid and vapour
bulks and on the wettability of the nanotube wall, we can
forecast when the fluid inside the nanotube is liquid or vapour; the wall
effect is dominant. The case of liquid and vapour separated by an interface
is not possible when the nanotube diameter is smaller than one hundred
nanometres and a liquid is generally found inside the nanotube. \newline
Recently, it was showed, by using nonequilibrium molecular dynamics
simulations, that liquid flow through a membrane composed of an array of
aligned carbon nanotubes is four to five orders of magnitude faster than it
would be predicted from conventional fluid-flow theory \cite{Majumder}.
These high fluid velocities are possible because of a frictionless
surface at the nanotube wall \cite{Ma}. Majunder \emph{et al} quote slip
lengths on the order of microns for their experiments with nanometer size
pores \cite{Majumder}. The  extremely large slip lengths measured in carbon
nanotubes greatly reduce the fluidic resistance  and
nanoscale structures could mimic extraordinarily fast flow possible in
biological cellular channels  \cite{Bonthuis}. By calculating the variation of water
viscosity and slip length as a function of the nanotubes diameter,   the results can be fully explained in the context of continuum
fluid mechanics \cite{Thomas}.\newline
In this paper we recalculate the  flows through nanotubes by using a
Navier-Stokes equation but, due to the slip condition and the Navier length,
it is possible to forecast an important difference between classical
Poiseuille flows and flows through nanotubes. The  calculations
associated with the physicochemical quality of the nanotube allow to forecast
if the fluid phase inside the tube is constituted of liquid or vapour. A  spectacular effect must appear when the mother bulk outside
the nanotube is constituted of vapour; in this case, the volume flow through the
nanotube is multiplied by a factor of the order of one million with respect to the Poiseuille flow and the
velocity field through the nanotube may be very important.

\section{Equation  of motion and boundary conditions for a capillary fluid}

\subsection{Case of conservative fluid}

The second gradient theory \cite{Forest,Germain}, conceptually more
straightforward than the Laplace theory, can be used to elaborate a theory
of capillarity \cite{Ono}. The theory can also be used to investigate
domains where the fluids are strongly inhomogeneous as in the immediate
vicinity of solid walls where intermolecular forces are dominant between
fluid and solid with respect to fluid interactions. By this simple way, the
only change with respect to compressible fluids is that the specific internal
energy is not only a function of the density $\rho $, of the specific
entropy $s$, but also of $\func{grad}\rho $. Consequently, the specific
internal energy $\varepsilon $ characterizes both the compressibility and
the surdeformation of the fluid,
\begin{equation*}
\varepsilon =f(\rho ,s,\beta ),\quad \mathrm{{where}\quad }\beta =(\func{grad%
}\rho )^{2}.
\end{equation*}
We recall the main results of \emph{capillary fluids} already obtained in
the literature \cite{Gouin 1,Gouin 4,Rowlinson}:

The equation of conservative motions of such \emph{capillary fluids} is
\begin{equation}
\rho \,\mathbf{a}=\func{div}\mathbf{\mathbf{\sigma }}-\rho\, \func{grad}%
\Omega ,  \label{2}
\end{equation}%
where  $\mathbf{a}$ denotes the acceleration vector, $\Omega $ the
extraneous force potential and $\mathbf{\mathbf{\sigma }}$ the total
stress tensor. The total stress tensor is
\begin{equation}
\mathbf{\mathbf{\sigma }}=-p\,\mathbf{I}-\lambda \,(\func{grad}\rho )(\func{%
grad}\rho )^{T}\ \ \mathrm{{or}\ }\ \sigma _{ij}=-p\ \delta _{ij}-\lambda
\,\rho _{,i}\rho _{,j}\,,\ i,j\in \{1,2,3\}  \label{3a}
\end{equation}%
where $\,^T$ denotes the transposition, with%
\begin{equation*}
\quad\lambda\equiv 2\,\rho \,\varepsilon _{\beta }^{\prime }\qquad \mathrm{{and}%
\qquad }p\equiv \rho ^{2}\varepsilon _{\rho }^{\prime }-\rho\,\func{div}(\lambda \,\func{%
grad}\rho ).  \label{sigma}
\end{equation*}%
It should be noted that $\varepsilon _{s}^{\prime }$ is the Kelvin
temperature expressed as a function of $\rho $, $s$ and $\beta $. It appears
that only the scalar $\lambda $ accounts for surdeformation effects. As $%
\varepsilon $ does, the scalar $\lambda$ depends on $\rho, s$ and $\beta $. For the
surface tension study based on the gas kinetic theory, Rocard obtained the
expression (\ref{3a}) for the stress tensor but with $\lambda$ constant \cite%
{Rocard}. If $\lambda $ is constant, the specific energy $\varepsilon $
reads
\begin{equation*}
\varepsilon (\rho ,s,\beta )=\alpha (\rho ,s)+\frac{\lambda}{2\rho }\ \beta ,
\end{equation*}%
and the \emph{second gradient} term $\displaystyle  {\lambda}\,
\beta/(2\,\rho)$ is simply added to the specific internal energy $\alpha
(\rho ,s)$ of the classical compressible fluid. The pressure of the
compressible fluid is $P \equiv \rho ^{2}\alpha _{\rho }^{\prime }$ and the
temperature is $T \equiv\alpha _{s}^{\prime }$. Consequently,
\begin{equation*}
p=P-\lambda \left(\frac{\beta }{2}+\rho\, \Delta \rho\right).
\end{equation*}%
For the thermodynamical pressure $P$, Rocard and others authors use the van der Waals
pressure
\begin{equation*}
P=\rho\, \frac{R\,T }{1-b\rho }-a\rho ^{2}
\end{equation*}%
or other similar laws  \cite{Rocard}. It should be
noted that if $\lambda$ is constant,  there exits a
relation independent of $\func{%
grad}\,\rho$ between $T ,\, \rho $ and $s$.

\subsubsection{Case $\lambda $ constant}

Equation (\ref{3a})  yields
\begin{equation*}
{\sigma }_{ij}=-P\,\delta _{ij}+\lambda \,\left\{ \left( \frac{1}{2}\ \rho
_{,k}\rho _{,k}+\rho \rho ,_{kk}\right) \delta _{ij}- \rho
_{,i}\rho _{,j}\right\}.
\end{equation*}
Let us denote $\omega =\Omega -\lambda \,\Delta \rho $, then Eq. (\ref{2}) reads
\begin{equation}
\rho \,\mathbf{a}+\func{grad}P+\rho \func{grad}\omega =0.  \label{motion2}
\end{equation}%
This relation is similar to the perfect fluid case; the term $\omega $
involves all capillarity effects. %
From ${\sigma }_{ij{,j}}=-P_{,i}+\lambda \,\rho \rho _{,ijj}\,  $ and by neglecting the extraneous forces, we obtain:
\begin{equation*}
\rho\,\mathbf{\mathbf{a}}+\func{grad}P=\lambda \,\rho \,\func{grad}%
\Delta \rho .
\end{equation*}

\subsubsection{Thermodynamic form of the equation of motion}

Commonly - and not only when $\lambda$ is constant - the equation of motion (%
\ref{2}) can be written in a thermodynamic form
\begin{equation}
\mathbf{a} =\theta \func{grad}s-\func{grad}(h+\Omega ) ,\quad \mathrm{with}%
\quad h= \varepsilon+\frac{p}{\rho}\, .  \label{thermotion}
\end{equation}
In the  non-capillarity case  ($\varepsilon ^{\prime }_{\beta } =0$ or $%
\varepsilon =\alpha (\rho ,s)$), Eq. (\ref{thermotion}) is well-known.
When $T$ is constant, Eq. (\ref{thermotion}) yields%
\begin{equation}
\mathbf{a} +\func{grad}(\pi+\Omega )=0 ,\quad {\rm with} \quad \pi=
h-T\,s\,.  \label{thermotion1}
\end{equation}
The potentials $h$ and $\pi$ are the \emph{generalized enthalpy and
the chemical potential} of the capillary fluid.

\subsection{Case of viscous fluid}

In the case of viscous fluids, the equation of motion includes not only the
stress tensor $\mathbf{\mathbf{\sigma }},$ but also the viscous stress
tensor $\mathbf{\mathbf{\sigma }}_{v}$ written in the form:
\begin{equation*}
\bm{\mathbf{\sigma }}_{v}=\eta \ \mathtt{tr}\,\bm{D}+2\,\kappa \,\bm{D},
\end{equation*}%
where $\bm D$ is the deformation tensor, symmetric gradient of the velocity
field and $\eta $ and $\kappa$ are constant in the viscous linear
case. Of course in second gradient theory, it would be coherent to add terms
accounting for the influence of higher order derivatives of the velocity
field to the viscous stress tensor $\mathbf{\mathbf{\sigma }}_{v}$; the
surdeformation of density comes from wall effects but the variations of velocity
 are negligible and the second
derivatives are not taken into account. Equation (\ref{2}) is modified by adding the forces associated with the
viscosity and we obtain
\begin{equation*}
\rho \,\mathbf{a}=\func{div}(\mathbf{\mathbf{\sigma }}+\mathbf{\mathbf{%
\sigma }}_{v})-\rho \func{grad}\,\Omega \,.
\end{equation*}%
For viscous fluid, Eq. (\ref{motion2}) reads
\begin{equation}
\rho \,\mathbf{a}+\func{grad}P+\rho \func{grad}(\Omega -\lambda \,\Delta
\rho )-\func{div}\mathbf{\mathbf{\sigma }}_{v}=0.  \label{viscous motions}
\end{equation}

\subsection{Boundary conditions at a solid wall}

The forces acting between liquid and solid   range over a few nanometres but can be simply described
 by  a special surface energy. This energy is not   the total interfacial energy which results from the direct fluid/solid contact;
  another
energy results from the distortion  in the fluid density profile near the wall.
For a solid wall  not too curved at a molecular scale, the total surface free energy $\varphi$ is developed as \cite%
{Gouin 2}:
\begin{equation}
\varphi (\rho_{_S})=-\gamma _{1}\rho_{_S} +\frac{1}{2}\,\gamma
_{2}\,\rho_{_S}^{2}.  \label{surface energy}
\end{equation}%
Here $\rho_{_S}$ denotes the limit value of the fluid density  at the
surface $(S)$; the constants $\gamma _{1}$, $\gamma _{2}$ as the constant $\lambda $ are
positive.  In the mean field approximation of molecular theory they are:
\begin{equation*}
\gamma _{1}=\frac{\pi c_{ls}}{12\delta ^{2}m_{l}m_{s}}\;\rho _{sol},\quad
\gamma _{2}=\frac{\pi c_{ll}}{12\delta ^{2}m_{l}^{2}},    \label{coefficients}
\end{equation*}
with
\begin{equation*}
\lambda =\frac{%
2\pi c_{ll}}{3\sigma _{l}\,m_{l}^{2}},
\end{equation*}
where $m_{l}$ et $m_{s}$ denote the molecular masses of fluid and
solid, respectively, $\rho _{sol}$ is the solid density; other constants come from London
potentials of liquid-liquid and liquid-solid interactions expressed in the
form
\begin{equation*}
\left\{
\begin{array}{c}
\displaystyle\;\;\;\;\;\;\varphi _{ll}=-\frac{c_{ll}}{r^{6}}\;,\text{ \
when\ }r>\sigma _{l}\;\;\text{and }\;\ \varphi _{ll}=\infty \text{ \ when \ }%
r\leq \sigma _{l}\,,\  \\
\displaystyle\;\;\;\;\;\;\varphi _{ls}=-\frac{c_{ls}}{r^{6}}\;,\text{ \
when\ }r>\delta \;\;\text{and }\;\ \varphi _{ls}=\infty \text{ \ when \ }%
r\leq \delta \;,\
\end{array}%
\right.
\end{equation*}%
where $c_{ll}$ et $c_{ls}$ are two positive coefficients associated with
Hamaker constants, $\sigma _{l}$ and $\sigma _{s}$ denote the molecular
diameters for the fluid and the solid, $\delta =\frac{1}{2}\,(\sigma _{l}+
\sigma _{s})$.

The boundary condition for the density at the solid wall $(S)$ associated with the free surface
energy (\ref{surface energy}) is calculated in   \cite{Gouin 3}:
\begin{equation}
\lambda \left( \frac{d\rho }{dn}\right) _{|_{S}}+\varphi ^{\prime
}(\rho_{_S})\ =0,  \label{cl1}
\end{equation}%
where $n\,$ is the external normal direction to the fluid.

\section{The chemical potential  in liquid and vapour phases}

The chemical potential of a compressible fluid at temperature $T$ is denoted
by $\mu _{_{0}}$. Due to the equation of state $P\equiv P(\rho ,T)$, it is
possible to express $\mu _{_{0}}$ as a function of $\rho $ (and $T$). At a
given temperature, the volume free energy $g_{_{0}}$ associated with $\mu _{_{0}}$
verifies $g_{_{0}}^{\prime }(\rho )=\mu _{_{0}}(\rho )$. Due to the fact $%
\mu _{_{0}}$ and $g_{_{0}}$ are defined to an additive constant, we add the
conditions
\begin{equation*}
\mu _{_{0}}(\rho _{l})=\mu _{_{0}}(\rho _{v})=0\quad {\mathrm{and}\quad
g_{_{0}}(\rho _{l})=g_{_{0}}(\rho }_{v}{)}=0,
\end{equation*}%
where $\rho _{l}$ and $\rho _{v}$ are the fluid densities in the liquid
and vapour bulks corresponding to the plane liquid-vapour interface at
temperature $T$.\newline
The expressions of the two thermodynamical potentials $\mu _{_0}$ and $%
g_{_0} $ can be expended at the first order near the liquid and vapour bulks,
respectively
\begin{equation*}
\mu _{_{0}}(\rho )=\frac{c_{l}^{2}}{\rho _{l}}\left( \rho -\rho _{l}\right)
\qquad \mathrm{{and}\qquad }\mu _{_{0}}(\rho )=\frac{c_{v}^{2}}{\rho _{v}}%
\left( \rho -\rho _{v}\right) ,
\end{equation*}%
\begin{equation*}
g_{_{0}}(\rho )=\frac{c_{l}^{2}}{2\rho _{l}}\left( \rho -\rho _{l}\right)
^{2}\qquad {\mathrm{and}\qquad g_{_{0}}(\rho )=\frac{c_{v}^{2}}{2\rho _{v}}%
\left( \rho -\rho _{v}\right) ^{2}},
\end{equation*}%
where $c_{l}$ and $c_{v}$ are the isothermal sound velocities in the liquid
and vapour bulks  \cite{Gouin 6}. Consequently, at temperature $T$, it is possible to obtain
the connection between the liquid bulk of density $\rho _{l_b}$ and the vapour
bulk of density $\rho _{v_{b}}$ corresponding to curved interfaces (as for
spherical bubbles and droplets \cite{Isola1,Isola2}):  we call them the
mother bulk densities. These equilibria do not obey   the Maxwell rule \cite{Aifantis}, but
the values of the chemical potential in the two mother bulks are equal:
\begin{equation}
\mu _{_{0}}(\rho _{l_b})=\mu _{_{0}}(\rho _{v_{b}}).  \label{equcp}
\end{equation}%
Consequently, we define $\mu _{l_{b}}(\rho )$ and $\mu _{v_{b}}(\rho )$ at
temperature $T$ as:
\begin{equation*}
\mu _{l_{b}}(\rho )=\mu _{_{0}}(\rho )-\mu _{_{0}}(\rho _{l_b})\equiv \mu
_{_{0}}(\rho )-\mu _{_{0}}(\rho _{v_{b}})=\mu _{v_{b}}(\rho ).
\end{equation*}
An expansion to the first order near the liquid and vapour bulks yields
\begin{equation*}
\mu _{l_{b}}(\rho )=\frac{c_{l}^{2}}{\rho _{l}}\left( \rho -\rho _{l_b}\right)
\qquad \mathrm{{and}\qquad }\mu _{v_{b}}(\rho )=\frac{c_{v}^{2}}{\rho _{v}}%
\left( \rho -\rho _{v_{b}}\right)
\end{equation*}%
and due to relation (\ref{equcp}),
\begin{equation*}
\frac{c_{l}^{2}}{\rho _{l}}\left( \rho _{l_b}-\rho _{l}\right) =\frac{c_{v}^{2}%
}{\rho _{v}}\left( \rho _{v_{b}}-\rho _{v}\right)
\end{equation*}%
which clarifies the connection   between $\rho _{l_b}$ and $\rho _{v_{b}}$.%
\newline
To the chemical potential $\mu _{l_{b}}(\rho)\equiv \mu _{v_{b}}(\rho)$ at temperature $T$, we
associate the volume free energies  $g_{l_{b}}(\rho )$ and
$g_{_{_{v_{b}}}}(\rho )$ that are null for $\rho _{l_b}$ and $\rho _{v_{b}}$,
respectively:
\begin{equation*}
g_{l_{b}}(\rho )=g_{_{0}}(\rho )-g_{_{0}}(\rho _{l_b})-\mu _{_{0}}(\rho _{l_b})(\rho -\rho _{l_b}),
\end{equation*}%
\begin{equation*}
g_{_{_{v_{b}}}}(\rho )=g_{_{0}}(\rho )-g_{_{0}}(\rho _{v_{b}})-\mu
_{_{0}}(\rho _{v_{b}})(\rho -\rho _{v_{b}}).
\end{equation*}
The free energies $g_{l_{b}}(\rho )$ and $g_{_{v_{b}}}(\rho )$ are  the
reference free energies associated with the liquid and vapour mother bulks. The reference free energy $g_{l_{b}}(\rho )$ differs from $%
g_{_{v_{b}}}(\rho )$ by a constant. Moreover, the volume free energies are
expanded as
\begin{equation*}
g_{l_{b}}(\rho )=\frac{c_{l}^{2}}{2\rho _{l}}\left( \rho -\rho _{l_b}\right)
^{2}\qquad \mathrm{{and}\qquad }g_{v_{b}}(\rho )=\frac{c_{v}^{2}}{2\rho _{v}}%
\left( \rho -\rho _{v_{b}}\right) ^{2}.
\end{equation*}

\section{Liquid and vapour densities in a nanotube}

A nanotube is constituted of a hollow cylinder of length size $\ell $ and of
small diameter $d=2R$, ($d/\ell \ll 1$). We consider solid walls with a
large thickness with regards to molecular dimensions such that the surface
energy verifies an expression in form (\ref{surface energy}). We assume that
a capillary fluid is a convenient model to represent fluids inside the
nanotube. \newline
At equilibrium, far from the nanotube tips and by neglecting the external forces, the profile of density is
solution of Eq. (\ref{thermotion1}) with $\mathbf{a}=0$ and $\pi =\mu
_{_{0}}-\lambda \,\Delta \rho $ :
\begin{equation}
\lambda \,\Delta \rho =\mu _{_{0}}(\rho )-C,  \label{densityequ}
\end{equation}%
where $C$ is an additional constant. The value of  $ C$ is associated with the
density value in the mother bulk outside  the nanotube
(where $\Delta \rho =0$)  \cite{Derjaguin}. Consequently, the reference density value $\rho
_{_{ref}}$ may be chosen as $\rho _{l_b}$ or $\rho _{v_{b}}$.\newline
We consider the cases when  exclusively liquid or vapour fill up the nanotube;
 the mother bulk can be as well liquid as vapour. The profile of density is
given by the differential equation:
\begin{equation}
\lambda \,\left( \frac{d^{2}u}{dr^{2}}+\frac{1}{r}\frac{du}{dr}\right) -%
\frac{c_{_{ref}}^{2}}{\rho _{_{ref}}}\ u=0,\qquad \mathrm{with}\quad u=\rho
-\rho _{_{ref}},  \label{densityprofile1}
\end{equation}%
where $c_{_{ref}}$ is the isothermal sound velocity associated with liquid
mother bulk (respectively vapour mother bulk) when the liquid phase fills up
the nanotube (respectively vapour phase). In cylindrical coordinates, $r$
denotes the radial coordinate. The reference length is
\begin{equation*}
\delta _{_{ref}}=\sqrt{{\,\lambda \,\rho _{_{ref}}}/{c_{_{ref}}^{2}}}\,.
\end{equation*}%
We denote by $x$ the  dimensionless variable such that $r=\delta
_{_{ref}}\,x$. Equation (\ref{densityprofile1}) reads
\begin{equation*}
\frac{d^{2}u}{dx^{2}}+\frac{1}{x}\frac{du}{dx}-\ u=0.
\label{densityprofile2}
\end{equation*}%
The solutions of Eq. (\ref{densityprofile2}) in classical
expansion form $u=\sum_{n=0}^{\infty }a_{n}x^{n}$  yield:
\begin{equation*}
\sum_{n=2}^{\infty }n^{2}\,a_{n}\,x^{n-2}-a_{n-2}\,x^{n-2}=0\quad
\Longrightarrow \quad n^{2}\,a_{n}=a_{n-2}\,.
\end{equation*}%
Due to the symmetry at $x=0$, the odd terms are null and consequently,
\begin{equation*}
u=a_{_{0}}\,\sum_{p=0}^{\infty }\ \frac{1}{4^{p}\,(p\,!)^{2}}\ x^{2p}\,.
\end{equation*}%
The series has an infinite radius of convergence. Let us define the
quantities
\begin{eqnarray*}
\qquad \qquad f(x) &=&\sum_{p=0}^{\infty }\ \frac{1}{4^{p}\,(p!)^{2}}\ x^{2p},
\\
\qquad \qquad h(x) &\equiv &f^{\prime }(x)=\sum_{p=1}^{\infty }\ \frac{2p}{%
4^{p}\,(p!)^{2}}\ x^{2p-1}, \\
\qquad \qquad k(x) &\equiv &f^{\prime \prime }(x)=\sum_{p=1}^{\infty }\ \frac{%
2p\,(2p-1)}{4^{p}\,(p!)^{2}}\ x^{2p-2}.
\end{eqnarray*}%
Consequently, $u=a_{_{0}}\,f(r/\delta _{_{ref}})$. The boundary condition (%
\ref{cl1}) at $x=R/\delta _{_{ref}}$ is:
\begin{equation*}
\lambda \,\frac{du}{dx}=  \gamma _{1}-\gamma _{2}\,\rho
  \qquad {\rm or} \qquad a_{_{0}}=\frac{\delta
_{_{ref}}\,\left( \gamma _{1}-\gamma _{2}\,\rho _{_{ref}}\right) }{\lambda
\,h\left( \frac{R}{\delta _{_{ref}}}\right) +\gamma _{2}\,\delta
_{_{ref}}\,f\left( \frac{R}{\delta _{_{ref}}}\right) }
\end{equation*}%
and the density profile reads
\begin{equation*}
\rho =\rho _{_{ref}}+\frac{\delta _{_{ref}}\,\left( \gamma _{1}-\gamma
_{2}\,\rho _{_{ref}}\right) }{\lambda \,h\left( \frac{R}{\delta _{_{ref}}}%
\right) +\gamma _{2}\,\delta _{_{ref}}\,f\left( \frac{R}{\delta _{_{ref}}}%
\right) }\ f\left( \frac{r}{\delta _{_{ref}}}\right) .
\end{equation*}%
Let us consider the free volume energy $g_{\rho _{ref}}(\rho )$, null in the
mother bulk of density $\rho _{_{ref}}$ (where $\rho _{_{ref}}=\rho _{l_b}$ or
$\rho _{_{ref}}=\rho _{v_{b}}$) chosen as the reference mother bulk; the
free energy $\phi $ per unit of volume in a inhomogeneous fluid is
\begin{equation*}
\phi =g_{\rho _{ref}}(\rho )+\frac{\lambda }{2}\,\left( \frac{d\rho }{dr}%
\right) ^{2}
\end{equation*}%
and consequently, if $\phi $ is expressed as a function of $r$,
\begin{eqnarray}
\phi (r) &=&g_{\rho _{ref}}\left( \rho _{_{ref}}+\frac{\delta
_{_{ref}}\,\left( \gamma _{1}-\gamma _{2}\,\rho _{_{ref}}\right) f\left(
\frac{r}{\delta _{_{ref}}}\right) }{\lambda \,h\left( \frac{R}{\delta
_{_{ref}}}\right) +\gamma _{2}\,\delta _{_{ref}}f\left( \frac{R}{\delta
_{_{ref}}}\right) }\right) + \notag \\
&&\,\frac{\lambda }{2}\left( \frac{\left( \gamma _{1}-\gamma _{2}\,\rho
_{_{ref}}\right) \,h\left( \frac{r}{\delta _{_{ref}}}\right) }{\lambda
\,h\left( \frac{R}{\delta _{_{ref}}}\right) +\gamma _{2}\,\delta
_{_{ref}}\,f\left( \frac{R}{\delta _{_{ref}}}\right) }\right) ^{2}\ .\label{freenergy}
\end{eqnarray}

\subsection{Impossibility of a two-phase fluid in a nanotube}
\begin{figure}[h]
\begin{center}
\includegraphics[width=6cm]{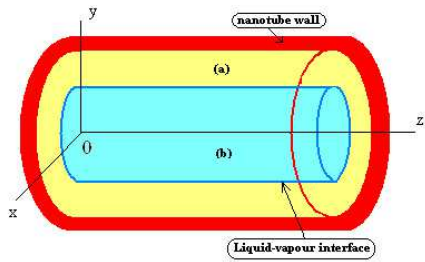}
\end{center}
\caption{\textbf{Two-phase fluid in a nanotube}: The nanotube is
simultaneously filled with two phases liquid and vapour of the same fluid. The two phases (a)
and (b) are separated by a cylindric \emph{material} interface.}
\label{Fig. 1}
\end{figure}

Let us consider a nanotube simultaneously filled with
liquid and vapour of the same fluid; an interface appears between the liquid
and vapour phases. By reason of  symmetry  the liquid-vapour interface is a material surface represented
by a cylindrical surface with the same axis as  the nanotube. The interface has a
positive surface energy $\gamma _{lv}$ increasing the free energy of the
fluid inside the nanotube. \\

\emph{First case:} \emph{domain} (a) \emph{is liquid and  domain} (b) \emph{is vapour.} An approximation allows us to compare the
energy of the only  liquid phase and the energy of the two-phase fluid when the
liquid is in contact with the nanotube wall. The free energies associated
with the wall are approximatively equal. We have the relation  $g_{l_{b}}(\rho )\equiv g_{v_{b}}(\rho )+g_{l_{b}}(\rho
_{v_{b}})$ with $- g_{l_{b}}(\rho _{v_{b}})=P(\rho _{v_b})-P(\rho _{l_{b}})= \Pi
(\rho _{l_b})$;  the term $\Pi (\rho _{l_b})$ is called \emph{the disjoining pressure} relatively
to the mother bulk $\rho_{l_b}$ \cite{Derjaguin,Gouin 6}.

The
difference between  the free energy per unit length $E_1$ of the liquid phase and the
free energy per unit length $E_2$ of the two-phase fluid is approximatively
\begin{equation*}
E_2-E_1=\pi \left( -e_{1}^{2}\,\Pi (\rho _{l_b})+2e_{1}\gamma _{lv}\right) ,
\end{equation*}%
where $e_{1}$ denotes the radius of the domain (b) of vapour
delimited by the interface. Consequently, the free energy for the two-phase fluid in the nanotube is
smaller  than for only the liquid phase if $%
e_{1}\geq 2\gamma _{lv}/\Pi (\rho _{l_b})$.\newline

\emph{Second case:} \emph{domain} (a) \emph{is vapour and domain} (b) \emph{is liquid.} We can also compare the energy of the only  vapour phase and the energy  of
the two-phase fluid when the vapour is in contact with the nanotube wall.
The free energies associated with the wall are approximatively equal.

The difference between the free energy per unit
length $E_3$ of the vapour phase and the free energy per unit length $E_4$ of
the two-phase fluid is approximatively
\begin{equation*}
E_4-E_3=\pi \left( e_{2}^{2}\,\Pi (\rho _{l_b})+2e_{2}\gamma _{lv}\right) ,
\end{equation*}%
where $e_{2}$ denotes the radius of the domain (b) of liquid
delimited by the interface. Consequently, the free energy in the nanotube is
smaller for the two-phase fluid than for the vapour phase if $%
e_{2}\geq 2\gamma _{lv}/(-\Pi (\rho _{l_b})$.\newline

As an example, we consider the case of water at $20^{\degree}$ Celsius, in
\textbf{c.g.s.} units, the interfacial free energy $\gamma _{lv}=72$. If $%
|\Pi (\rho _{l_b})|=10^{7}$ (or 10 atmospheres), corresponding to an important
absolute value of the disjoining pressure, we obtain $e_{1}=e_{2}\geq R_0 =
14.4\times 10^{-6}=144$\thinspace\ nm$\, = 0.144\,\mu $m. Consequently, the nanotube is filled with only one phase if its radius
verifies the inequality $R < R_0$. The limit radius $R_0$ corresponds to the radius of  a
microscopic tube.  Consequently, we have just to compare the
free energies of the liquid  and  the vapour phases filling up the nanotube.

\subsection{Liquid phase in the nanotube}

 We consider the case when the fluid phase in the nanotube is
 liquid; the liquid density is close to $\rho _{l_b}$ (and $\rho _{l_b}\simeq
\rho _{l}$). We choose   $g_{l_{b}}$ as \emph{reference level of volume free energy}.
By taking account of Eq. (\ref{freenergy}) and  $\delta _{l}=\sqrt{%
\lambda \rho _{l}/c_{l}^{2}}$, we get
\begin{equation}
\phi(r)=\frac{c_{l}^{2}}{2\rho _{l}}\left( \frac{\delta _{l}\,\left( \gamma
_{1}-\gamma _{2}\,\rho _{l_b}\right) f\left( \frac{r}{\delta _{l}}\right) }{%
\lambda \,h\left( \frac{R}{\delta _{l}}\right) +\gamma _{2}\,\delta
_{l}f\left( \frac{R}{\delta _{l}}\right) }\right) ^{2}+ \frac{\lambda }{2}%
\left( \frac{\left( \gamma _{1}-\gamma _{2}\,\rho _{l_b}\right)  h\left(
\frac{r}{\delta _{l}}\right) }{\lambda \,h\left( \frac{R}{\delta _{l}}%
\right) +\gamma _{2}\,\delta _{l}\,f\left( \frac{R}{\delta _{l}}\right) }%
\right) ^{2}.  \label{freenergy1}
\end{equation}%
Due to the fact that $c_{l}^{2}\delta _{l}^{2}/2\rho _{l}=\lambda /2$, Eq. (%
\ref{freenergy1}) reads
\begin{equation*}
\phi (r)=\frac{\lambda }{2}\,\left( \gamma _{1}-\gamma _{2}\,\rho _{l_b}\right) ^{2}\frac{f\left( \frac{r}{\delta _{l}}\right) ^{2}+h\left(
\frac{r}{\delta _{l}}\right) ^{2}}{\left( \lambda \,h\left( \frac{R}{\delta
_{l}}\right) +\gamma _{2}\,\delta _{l}\,f\left( \frac{R}{\delta _{l}}\right)
\right) ^{2}}\ .
\end{equation*}%
The total free energy per unit of length in the nanotube is:
\begin{equation*}
E_{l_b}(\rho _{l_b})=2\pi \left( \int_{0}^{R}\phi (r)\,r\,dr+R\left( -\gamma
_{1}+\frac{\gamma _{2}}{2}\,\rho _{_{R}}\right) \rho _{_{R}}\right) ,
\end{equation*}%
where
\begin{equation*}
\rho _{_{R}}=\rho _{l_b}+\frac{\delta _{l}\,\left( \gamma _{1}-\gamma
_{2}\,\rho _{l_b}\right) }{\gamma _{2}\,\delta _{l}\,f\left( \frac{R}{\delta
_{l}}\right) +\lambda \,h\left( \frac{R}{\delta _{l}}\right) }\,f\left(
\frac{R}{\delta _{l}}\right) .
\end{equation*}%
Consequently, if we denote $r=\delta _{l}\,x$ and $n=R/\,\delta _{l}$, we
obtain the total free energy per unit of surface in the nanotube $F_{l_b}(\rho _{l_b})\equiv E_{l_b}(l)/2\pi R$ in the form
\begin{equation}
F_{l_b}(\rho _{l_b})=\frac{\delta _{l}c_{l}^{2}}{2\rho _{l}}\frac{\left( \gamma
_{1}-\gamma _{2}\,\rho _{l_b}\right) ^{2}}{n}{{\int_{0}^{n}}}\frac{f\left(
x\right) ^{2}+h\left( x\right) ^{2}}{\left( \gamma _{2}\,f\left( n\right) +%
\frac{\lambda }{\delta _{l}}\,h\left( n\right) \right) ^{2}}\,x\,dx+\left(
-\gamma _{1}+\frac{\gamma _{2}}{2}\,\rho _{_{R}}\right) \rho _{_{R}}.
\label{free energy1}
\end{equation}

\subsection{Vapour in the nanotube}

We consider the case when the fluid phase in the nanotube
is  vapour; the vapour density is close to $\rho _{v_{b}}$ (and $ \rho _{v_{b}}\simeq \rho _{v}$). For the reference level of volume
free energy we obtain for a density close to $\rho _{v}$
\begin{equation*}
g_{l_{b}}(\rho )={\frac{c_{v}^{2}}{2\rho _{v}}\left( \rho -\rho _{v}\right)
^{2}}-\Pi(\rho _{l_b}).
\end{equation*}%
The density in the
nanotube is close to the vapour density $\rho _{v_{b}}$ and consequently  we   neglect
the surface free energy of the wall. With $%
\gamma _{1}-\gamma _{2}\,\rho _{v_{b}}\approx \gamma _{1}$ we get
\begin{equation*}
\phi (r)\approx \psi (r)-\Pi (\rho _{l_b})\ \text{\ with}\ \ \ \psi (r)=\frac{%
\lambda }{2}\,\gamma _{1}^{2}\ \frac{f\left( \frac{r}{\delta _{v}}\right)
^{2}+h\left( \frac{r}{\delta _{v}}\right) ^{2}}{\left( \lambda \,h\left(
\frac{R}{\delta _{v}}\right) +\gamma _{2}\,\delta _{v}\,f\left( \frac{R}{%
\delta _{v}}\right) \right) ^{2}}\ ,
\end{equation*}
where   $\delta _{v}=\sqrt{\lambda \rho _{v}/c_{v}^{2}}$.
The total free energy per unit of length in the nanotube is:
\begin{equation*}
E_{l_b}(\rho {_{v_{b}}})=2\pi \left( \int_{0}^{R}\psi (r)\,r\,dr-\Pi (\rho _{l_b})%
\frac{R^{2}}{2}\right).
\end{equation*}%
If we denote $r=\delta _{v}\,x$ and $N=R/\delta _{v}$, we obtain the total
free energy per unit of surface in the nanotube $F_{l_b}(\rho _{v_{b}})\equiv
E_{l_b}(v)/2\, \pi\, R$ in the form
\begin{equation}
F_{l_b}(\rho {_{v_{b}}})=\frac{\delta _{v}\,c_{v}^{2}}{2\rho _{v}}\frac{\gamma
_{1}^{2}}{N}{{\int_{0}^{N}}}\frac{f\left( x\right) ^{2}+g\left( x\right) ^{2}%
}{\left( \gamma _{2}\,f\left( N\right) +\frac{\lambda }{\delta _{v}}
\,g\left( N\right) \right) ^{2}}\,x\,dx\ -\frac{N\,\delta _{v}}{2}\,\Pi
(\rho _{l_b})\, . \label{free energy2}
\end{equation}

\subsection{Numerical application in the case of water}

\begin{figure}[h]
\begin{center}
\includegraphics[width=9cm]{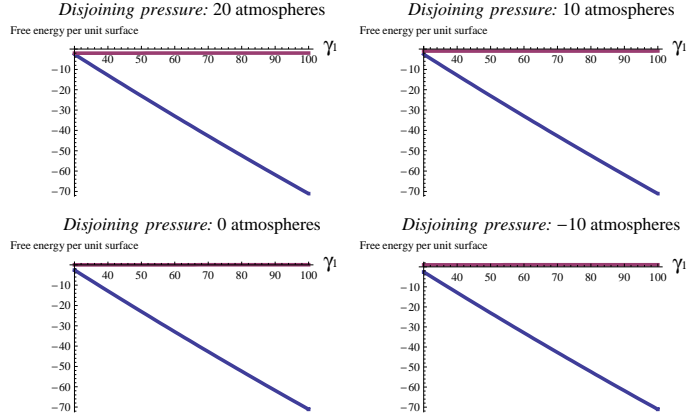}
\end{center}
\caption{\textbf{Nanotube of diameter 4 nm}: When we change the value of the
disjoining pressure $\Pi(\protect\rho _{l_b})$, the total free energy in a
nanotube filled up with liquid water phase is smaller than the total free
energy of a nanotube filled up with vapour water phase. Consequently, the
nanotube is always filled up with a liquid water phase.}
\label{Fig. 2}
\end{figure}
\begin{figure}[h]
\begin{center}
\includegraphics[width=9cm]{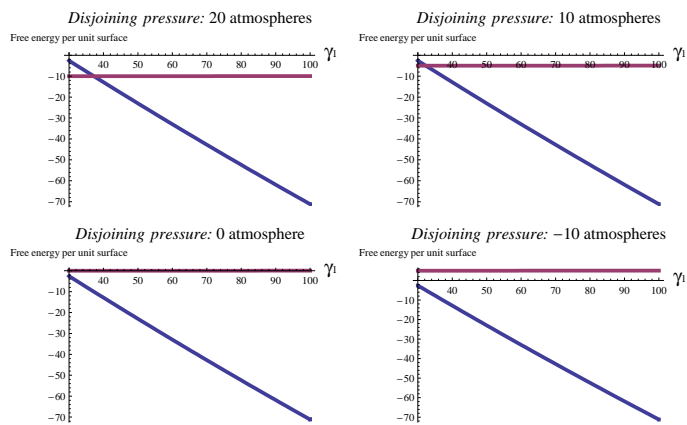}
\end{center}
\caption{\textbf{Nanotube of diameter 20 nm}: When the disjoining pressure
is positive and strong enough and if the wall is strongly hydrophobic, the
nanotube can be filled up with a water vapour phase. For an hydrophillic wall the
nanotube is always filled up with a liquid water phase.}
\label{Fig. 3}
\end{figure}

\begin{figure}[h]
\begin{center}
\includegraphics[width=9cm]{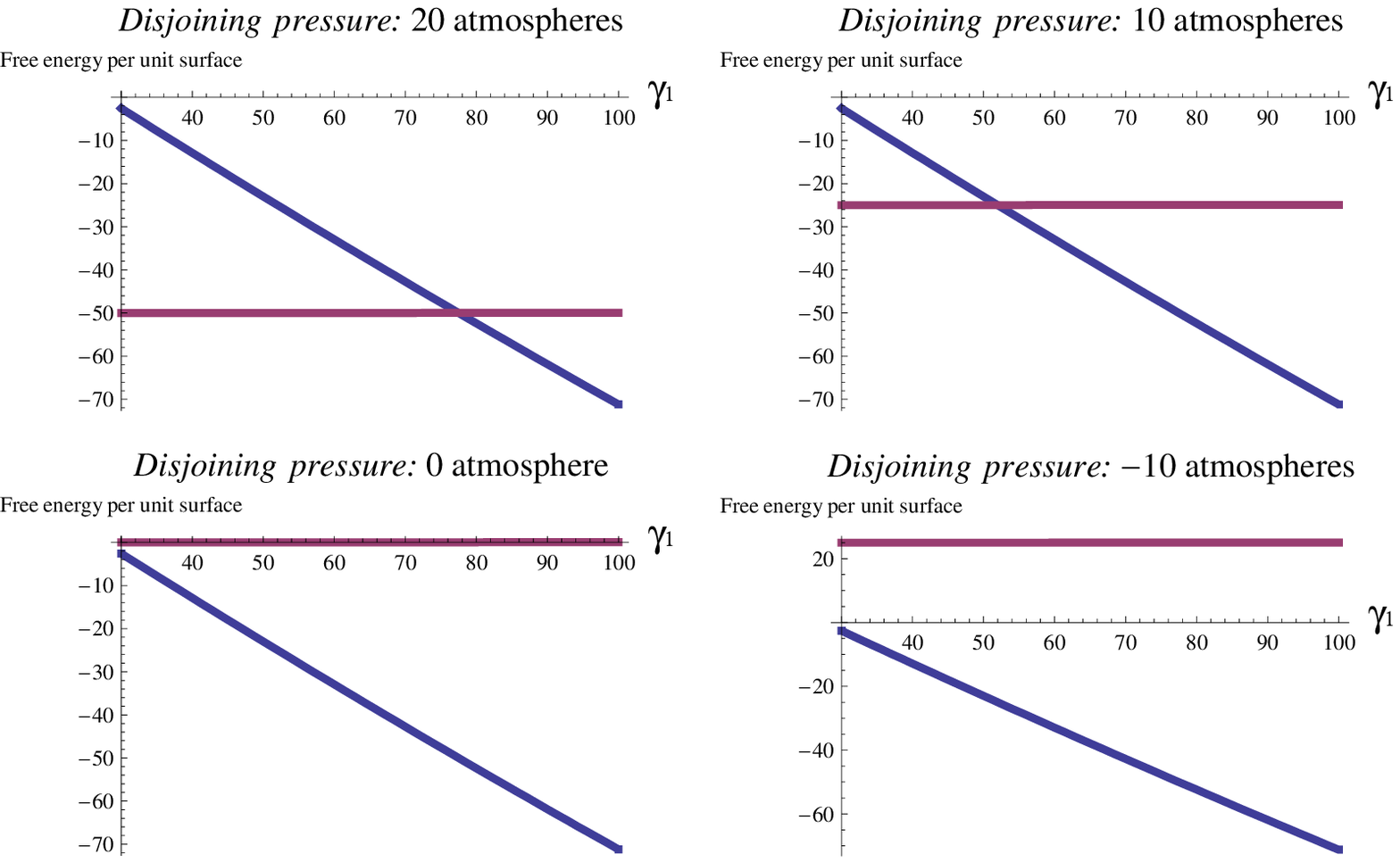}
\end{center}
\caption{\textbf{Nanotube of diameter 100 nm}: The tube corresponding to a
diameter of about $0.1 \,\mu$m is  a microtube. The quality of
the wall and the effect of the disjoining pressure are in competition. If
the disjoining pressure is strong enough and if the wall is moderately
hydrophobic, the tube is filled up with vapour water phase. When the
disjoining pressure is negative, the tube is always filled up with liquid water
phase.}
\label{Fig. 4}
\end{figure}

In \textbf{c.g.s.} units the physical constants of water are\ $c_{v}=3.7\
\times 10^{4};\ c_{l}=1.478\ \times 10^{5};\ \rho _{v}=1.7\ \times 10^{-6};
\rho _{l}=0.998;\lambda =1.17\ \times 10^{-5};\ \gamma _{2}=54.$\newline
We have obtained the free energy values for liquid and vapour phases.
Relations (\ref{free energy1}-\ref{free energy2})  depend on both  the wetting
quality of the wall and on the value of the disjoining pressure. For
convenient materials, we can  numerically compare the free energy of the
liquid with the free energy of the vapour. We consider the case when the water
fluid is in contact with different nanotube walls. The $x$-axis corresponds to
the value of $\gamma_1$ associated with the hydrophobicity or
hydrophillicity of the wall. The value of $\gamma_2$ depends only on the
fluid. When $\gamma_1$ is small enough, the wall of the nanotube is
hydrophobic and when $\gamma_1$ is large enough, the wall of the nanotube is
hydrophillic. The $y$-axis corresponds to the value of the total free energy
in the nanotube per unit surface of the wall. The case when the nanotube
wall is strongly hydrophobic corresponds to $\gamma_1 < 30$.   In all the graphs the straight line parallel to the $x$%
-axis corresponds to the free energy per unit surface of the nanotube filled up
with a vapour phase; the oblique curve corresponds to  the free energy per
unit surface of the nanotube filled up with a liquid phase. The graphs
corresponding to the values of the two free energies (\ref{free energy1}-\ref{free energy2}) allow to foresee if the nanotube is filled up with liquid
or with vapour. They are presented in Figures 2 to 4.\newline
 When the disjoining pressure is negative, the
nanotube is  filled up with a liquid water phase for all diameters of the nanotube. When the nanotube wall is
hydrophobic, for a large radius and a strong positive disjoining pressure,
the nanotube can be filled up with a vapour phase. As a result, the case of  vapour phase
filling up the nanotube is  less  usual than the case of liquid
water phase.\newline
In all the cases, we note that the volume free energy of the phase in the
nanotube is negligible with respect to the surface free energy of the wall.

\section{Permanent viscous motions in a nanotube}
"Fluid flow through nanoscopic structures, such as carbon nanotubes, is very different from the corresponding flow through microscopic and macroscopic structures. For example, the flow of fluids through nanomachines is expected to be fundamentally different from the flow through large-scale machines since, for the latter flow, the atomistic degrees of freedom of the fluids can be safely ignored, and the flow in such structures can be characterized by viscosity, density and other bulk properties. Furthermore, for flows through large-scale systems, the no-slip boundary condition is often implemented, according to which the fluid velocity is negligibly small at the fluid/wall boundary. Reducing the length scales immediately introduces new phenomena into the physics of the problem, in addition to the fact that at nanoscopic scales the motion of both the walls and the fluid, and their mutual interaction, must be taken into account" \cite{Rafii}.

In this section, we consider the permanent and laminar motions of viscous capillary liquid in
a nanotube. Because the liquid is heterogeneous, for capillary fluids, the
liquid stress tensor is not anymore scalar and the equations of
hydrodynamics are not valid. However, the results obtained for viscous flows \cite{Landau}  can be adapted at nanoscale. \newline
As in \cite{Rocard}, we assume that the kinematic viscosity coefficient $\nu
=\kappa /\rho $ only depends on the temperature. In the equation of motions, the
viscosity term is
\begin{equation*}
\ ({1}/{\rho })\,\text{div }\mathbf{\sigma }_{v}=2\nu \,\left[ \ \text{div }{%
\bm D}\,+\,{\bm D}\text{ grad \{\thinspace Ln}\,(2\,\kappa )\}\ \right] ,
\end{equation*}%
where ${\bm D}$ is the velocity deformation tensor and ${\bm D}$ grad\{Ln ($%
2\,\kappa $)\} is negligible with respect to div\thinspace ${\bm D}$.
\newline
We denote the velocity by $\mathbf{V}=(0,0,w)^{T}$ where $w$ is the velocity
component in   direction of the nanotube axis. When we neglect the external forces (as
gravity), the liquid nano-motion verifies Eq. (\ref{viscous motions})
written in the form
\begin{equation}
{\mathbf{a}}+\text{grad}[\,\mu _{o}(\rho )-\lambda \,\Delta \rho \,]=\nu
\,\Delta {\mathbf{V}}\ \ \mathrm{with}\ \ \Delta {\mathbf{V}}\simeq
\begin{bmatrix}
0,0,\Delta w%
\end{bmatrix} ^T
.  \label{motion0}
\end{equation}%
\newline
Simple fluids slip on a solid wall only at a molecular level \cite{Churaev}
and consequently, in classical conditions, the kinematic condition at solid
wall is the adherence condition $(z=0\;\Rightarrow \;w=0)$.
Recent papers in nonequilibrium molecular dynamic simulations of three
dimensional micro-Poiseuille flows in Knudsen regime reconsider
micro-channels: the influence of gravity force, surface roughness, surface
wetting condition and wall density are investigated. The results point out
that the no-slip condition can be observed for Knudsen flow when the surface
is rough. The roughness is a dominant parameter when the slip of fluid
is concerned. The surface wetting condition substantially influences the
velocity profiles \cite{Tabeling}.
But it is not the case for smooth surfaces. The relation between wall shear
stress and slip velocity is the key for characterizing the slip flow. With
water flowing through  hydrophobic thin capillaries, there are some qualitative
evidences for slippage \cite{Blake}. De Gennes said: "the results are
unexpected and stimulating and led us to think about unusual processes which
could take place near a wall. They are connected with the thickness of the
film when the thickness is of an order of the mean free path" \cite{De
Gennes}.\newline
When the free mean path $L$ is smaller than $d$, the Knudsen number  \emph{Kn} is smaller
than 1. That is the case for liquid where the mean free path $L$ is of the
same order than the molecular diameter. For example in the case of liquid
water,  \emph{Kn}  ranges between $0.5$ and $10^{-2}$ while the nanotube radius ranges between $1$ nm and $50$ nm. The adherence boundary condition at
a surface,  commonly employed with the Navier-Stokes equation
assuming a zero flow, is physically invalid and a slip regime occurs; the
boundary condition must be changed to take account of the  slippage
at the solid surfaces. \newline
For gases, the mean free path is of order of one hundred molecular diameters
and consequently the flow regime is only valid for large nanotubes. For
thin nanotubes the rarefied gas regime must be considered; but the
calculation in slip regime may give an idea of the change of flow with
respect to the Navier-Stokes regime also for gases. Nonetheless, we note that the vapour
phase in the tube occurs for large nanotube relevant from the
microfluidic case and the slip condition using continuum mechanics is realistic.
In fluid/wall slippage, the condition at solid wall writes
\begin{equation}
w=L_{s}\frac{\partial w}{\partial r}\qquad {\mathrm{ at}}\  \ r=R,  \label{slip}
\end{equation}%
where $L_{s}$ is the so-called Navier length \cite{Landau}. The Navier
length is expected to be independent of the thickness of the nanoflow and
may be as large as a few microns \cite{Tabeling}.

In the following, the dynamics of liquid nanoflows is studied in the case of
nonrough   nanotubes. Consequently,\newline
\emph{i)} The equation of motion  writes in the form (\ref{motion0}), \newline
\emph{ii)} The boundary conditions take account of the slip condition (\ref%
{slip}). \newline
When the liquid nanoflow thickness is small with respect to transverse
dimensions of the wall, it is possible to simplify Eq. (\ref{motion0}) which
governs the viscous flow; so, when $d\ll \ell $, \newline
\textit{iii) }We consider a laminar flow: the velocity component along the
wall is large with respect to the normal velocity component to the wall
which is negligible.\newline
\textit{iv) }For permanent motion, the equation of continuity reads:
\begin{equation*}
\left(\mathtt{grad}\,\rho\right)^T   \mathbf{V}+\rho _{\ }\mathtt{div}\,%
\mathbf{V}=0.
\end{equation*}%
The velocity vector $\mathbf{V}$ mainly varies along the direction
orthogonal to the wall and grad\thinspace $\rho $ is normal
to $\mathbf{V}$.
   The density is constant along each stream line ($\overset{\mathbf{%
\centerdot }}{\rho }=0\Longleftrightarrow \mathtt{div}\,\mathbf{V}=0$); the trajectories are drawn on isodensity surfaces
and $w=w(r)$. Due to the solid wall effect, the density in the tube is closely constant out from a boundary layer of approximatively one nanometer \cite{Gouin 6}; consequently, we consider the approximation of an incompressible liquid  in the tube.\newline
\emph{v)} Due to the geometry of the tube, for
permanent motion, the acceleration is null.
Equations (\ref{viscous motions}) or   (\ref{motion0}) separate  as:

$\bullet \ \ $ The first part along the $z$-coordinate,
\begin{equation}
\frac{\partial P}{\partial z}=\kappa \,\Delta \,w\qquad \mathrm{with}\qquad
\Delta \,w=\frac{1}{r}\frac{d}{dr}\left( r\ \frac{dw}{dr}\right),
\label{axequ}
\end{equation}
\indent $\bullet \ \ $ The second part in the plane orthogonal to the tube axis,
\begin{equation}
\frac{\partial }{\partial r}\,(\lambda \,\Delta \rho -\mu _{_{0}})=0.
\label{tgpart}
\end{equation}%
Equation (\ref{tgpart}) yields the same equation (\ref{densityequ}) as at
equilibrium. Equation (\ref{axequ}) yields
\begin{equation}
\frac{1}{r}\frac{d}{dr}\left( r\ \frac{dw}{dr}\right) =-\frac{\wp }{\kappa}\,,
\label{axequ1}
\end{equation}%
where $\wp $ denotes the pressure gradient along the nanotube. The cylindrical
symmetry of the nanotube yields the solution of Eq. (\ref{axequ1}) in the
form
\begin{equation*}
w=-\frac{\wp }{\kappa }\,\frac{r^{2}}{4}+b,
\end{equation*}
where $b$ is constant.
Condition (\ref{slip}) implies
\begin{equation*}
-\frac{\wp }{4\kappa}R^{2}+b=L_{s}\frac{\wp }{2\kappa }R
\end{equation*}%
and consequently,
\begin{equation*}
w=\frac{\wp }{4\kappa }\left( -r^{2}+R(R+L_{s})\right).
\end{equation*}%
The density in the nanotube is closely equal to $\rho _{l}$ and the volume
flow through the nanotube is $Q=2\pi \displaystyle\int_{0}^{R}w\,r\,dr=\frac{\pi\,\wp }{%
8\,\kappa }\,R^{3}(R+4L_{s})$. With  $Q_{o}$ denoting the Poiseuille flow
corresponding to a tube of the same radius $R$, we obtain:
\begin{equation}
Q=Q_{o}\left( 1+\frac{4L_{s}}{R}\right).  \label{flow}
\end{equation}%
In most cases, the Navier length is   of the micron order  ($%
L_{s}=1\,\mu m=10^{3}\,$nm) \cite{Majumder}. If we consider a nanotube with   $R=2\,$%
nm, we obtain $Q=2\times 10^{3}\,Q_{o}$. For $R=50\,$nm, that we consider as the
maximum radius of nanotube with respect to microfluidics, we
obtain $Q=40\,Q_{o}$. Consequently, the flow of liquid in nanofluidics is dramatically more
important than the Poiseuille flow in cylindrical tubes. In the
case of gases, we obtain the same results for nanotube of radius $R=50\,$nm
corresponding to a Knudsen number smaller than $0.5$. For nanotubes of radius
smaller than $30\,$nm, when the nanotube is unusually filled up with a vapour
phase, the flow is not anymore a continuous flow but is relevant  to kinetic
of rarified gases. Nevertheless, the magnitude of this flow is of several order more important than   Poiseuille flow.\newline
We have to emphasis that, when the mother bulk is vapour, in classical Poiseuille flow, the phase is vapour in the tube  but, for a nanotube  the phase is generally liquid (as in conditions presented in Fig. 2 and Fig. 3) and the volume flow through the nanotube is approximatively:
\begin{equation*}
Q=Q_{o}\frac{\rho _{l}}{\rho _{v}}\left( 1+\frac{4L_{s}}{R}\right).
\label{flow2}
\end{equation*}%
In the case of water  $\rho _{l}/ \rho _{v}\sim 10^3$, we get a
volume flow $10^{3}$ time more important than the volume  flow obtained in Eq. (\ref%
{flow}):
\begin{equation*}
Q \sim 10^{6}\,Q_{o}\,.
\end{equation*}%

\section{Conclusion}

A nanotube with a diameter ranging between 4 and 100 nanometres  is filled up only with one liquid phase or one vapour phase independently of the external mother bulk. For nanotubes with diameters smaller than 20 nm, the fluid phase inside the nanotube is generally liquid. For nanotubes of large diameters with respect to the molecular scale, the fluid phase can be liquid or vapour according to the values of the disjoining pressure  and of the physicochemical properties of the tube walls.\newline For nanotubes with small diameters, the flows can be significantly greater than usual  Poiseuille flows, especially if the mother bulk consists of vapour.\newline These results, obtained by using a nonlinear model of continuum mechanics and its associated differential equations, are in good agreement with experiments and molecular dynamics calculations.
\\

\noindent \textbf{Acknowledgements:} The paper has been supported by the XVI Conference  on Waves and Stability in Continuous Media and Italian \emph{Gruppo Nazionale per la Fisica Matematica}.

\end{document}